\documentclass{PoS}

\newcommand{\msub}[1]{\ensuremath _{\mbox{\scriptsize #1}}}

\title{Gapless Dirac Spectrum at High Temperature}

\ShortTitle{Gapless Dirac Spectrum at High Temperature}

\author{\speaker{Tamas G. Kovacs}\thanks{Research supported by OTKA Grant 
No.\ T046925. I thank Poul Damgaard for helpful correspondence on random
matrix theory.}\\
        Department of Physics, University of P\'ecs \\
        H-7624 P\'ecs, Ifj\'us\'ag \'utja 6. \\
        E-mail: \email{kgt@fizika.ttk.pte.hu}}

\abstract{
Using the overlap Dirac operator I show that, contrary to some expectations,
even well above the critical temperature there is not necessarily a gap in the
Dirac spectrum in pure SU(2) gauge theory. This happens when the Polyakov loop
and the fermion boundary condition combine to give close to periodic boundary
condition for the fermions in the time direction.  In this Polyakov loop
sector there is a non-vanishing density of Dirac eigenvalues around zero
which implies that chiral symmetry is spontaneously broken. I
demonstrate this both directly and also by finding good agreement with the 
random matrix theory prediction for the distribution of the lowest Dirac 
eigenvalue. I show that the chiral condensate increases with the temperature
therefore it is very unlikely to be explained by topological fluctuations
that become rapidly smaller above $T_c$. Finally I show that it is only a
small fraction of the lowest Dirac eigenvalues that decide which Polyakov
loop sector is favored by the fermion determinant if dynamical fermions 
are turned on. This provides a qualitative understanding of how the loss
of confinement above $T_c$ implies the restoration of chiral symmetry.
}

\FullConference{The XXVI International Symposium on Lattice Field Theory\\
		 July 14-19 2008\\
		 Williamsburg, Virginia, USA}

\begin{document}

\section{Introduction}

It is generally accepted that four dimensional $SU(N)$ gauge theories have a
high temperature deconfined phase with the Polyakov loop $Z(N)$ symmetry
spontaneously broken.  This is in the theory without dynamical fermions.
Including massless dynamical fermions will change this picture since the
fermionic determinant breaks the $Z(N)$ symmetry explicitly. However, even in
this case the transition from the low to the high temperature phase is
characterized by a substantial increase of the expectation value of the
Polyakov loop.

Another important feature of the theory with massless fermions is that the
transition to the high temperature phase is accompanied by the restoration of
chiral symmetry that is spontaneously broken at low temperatures. Generally
in QCD-like theories the chiral and the deconfinement transition take 
place at roughly the same temperature regardless of whether the transition is a 
cross-over or a genuine phase transition \cite{Aoki:2006br}.
How the deconfining and the chiral transition are linked is an important
and as yet even qualitatively not understood question. 

The study of the interplay between the Polyakov loop and chiral symmetry
restoration has a long history.  More than 10 years ago Chandrasekharan and
Christ noticed that in quenched QCD the chiral condensate vanishes in the high
temperature phase only if the phase of the Polyakov loop is real
\cite{Chandrasekharan:1995gt}. Inspired by this intriguing result Stephanov
used random matrix theory to predict that both in the $SU(2)$ and $SU(3)$ case
the chiral restoration temperature depends on the Polyakov loop
sector. Moreover, in the SU(2) case chiral symmetry is expected to remain
broken at arbitrarily high temperature provided the Polyakov loop is negative
\cite{Stephanov:1996he}. Further support to this scenario was given by
calculations in the Nambu-Jona-Lasinio model
\cite{Meisinger:1995ih,Chandrasekharan:1995nf}.

More recently direct lattice simulations have also been performed to check
whether this really happens.  Based on the appearance of a spectral gap above
$T_c$ in the $SU(3)$ case, Gattringer et al.\ concluded that the chiral
restoration occurs at the same temperature in all Polyakov loop sectors.
They, however, found that the spectral gap above $T_c$ depends on the Polyakov
loop sector \cite{Gattringer:2002dv}. This is also indicated by the behavior
of the so called dual quark condensate \cite{Bilgici:2008qy}. In the 
$SU(2)$ case, Bornyakov et al.\ concluded that in the negative Polyakov 
loop sector chiral symmetry remains broken up to $T=2T_c$
\cite{Bornyakov:2007xe,Bornyakov:2008bg}.

In summary, the somewhat controversial picture is that the Polyakov loop
has a strong influence on the chiral condensate, especially in the $SU(2)$
case. A further question in this connection is how a chiral condensate well 
in the high temperature phase can be understood based on mixing instanton
anti-instanton zero modes, given that the topological charge density rapidly
drops above $T_c$.

To shed some more light on these questions, in the present paper I study the
low end of the spectrum of the overlap Dirac operator in quenched $SU(2)$
gauge backgrounds well above $T_c$.  In particular I look at how the spectrum
is influenced by the Polyakov loop sector in the high temperature phase. I
also study how the fermion determinant ``selects'' a given Polyakov loop
sector and how this mechanism connects deconfinement and chiral symmetry
restoration.

\section{Simulation parameters}

Let us first summarize the parameters of the simulations. All of the
runs are quenched Wilson action $SU(2)$ lattices at $\beta=2.60$. This
is the critical $\beta$ for $N_T=10.4$. In the present study I chose to fix
$\beta$ to avoid renormalization issues for the chiral condensate. I varied
the temperature by choosing $N_T=4$ and $6$, which correspond to $T=2.6T_c$
and $1.7T_c$ respectively. I also varied the spatial size of the box to
check the scaling of the low eigenvalue density with the volume.   

On these lattices I computed the lowest 16-32 eigenvalues of the overlap Dirac
operator \cite{Narayanan:1994gw}, which is defined in terms of the Wilson
Dirac operator $D_w$ as
\begin{equation}
 D\msub{ov} = 1 - A \left[A^\dagger A\right]^{-\frac{1}{2}},
 \hspace{1cm} A = 1+s - D_w.
\end{equation}
The real parameter $s$ was chosen to be $0.4$ by maximizing the average lowest
eigenvalue of the kernel $A^\dagger A$.

\section{Chiral condensate}

\begin{figure}
\begin{center}
\includegraphics[width=.7\textwidth]{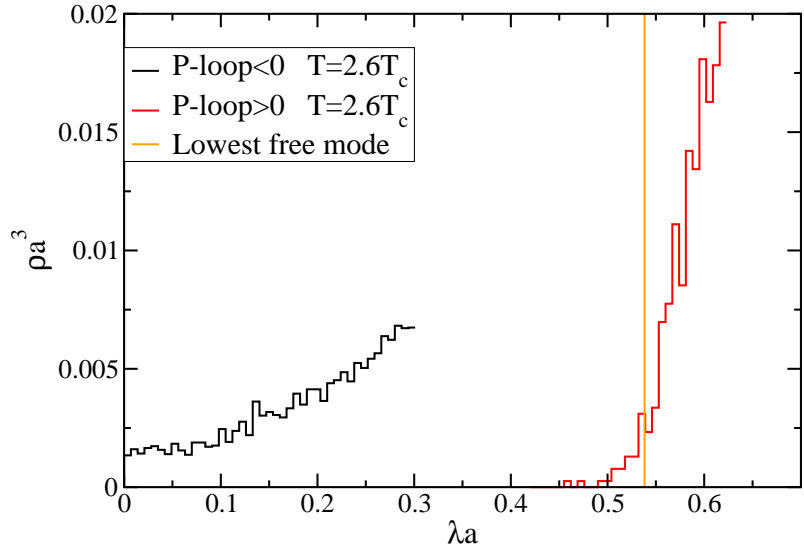}
\caption{The density of low eigenmodes of the overlap Dirac operator in 
the two Polyakov loop sectors.
The vertical line indicates the lowest Matsubara mode in the given 
box size.}
\label{fig:denslow_sch0}
\end{center}
\end{figure}

To see how the Polyakov loop influences the chiral condensate in
Figure \ref{fig:denslow_sch0} I plotted the low end of the eigenvalue density
of the Dirac operator in both Polyakov loop sectors. Here the temperature was
chosen to be $T=2.6T_c$ ($N_T=4$), well in the high temperature phase. The
difference between the two sectors is dramatic.
On the one hand, in the $P>0$ positive Polyakov loop sector there is a
sizeable gap in the spectrum and the density of modes at $\lambda=0$ is
clearly zero.  On the other hand, in the $P<0$ sector the eigenvalue density
at $\lambda=0$ is obviously non-vanishing. Through the Banks-Casher relation
\cite{Banks:1979yr} this implies that at this temperature chiral symmetry is
restored only if the Polyakov loop is positive.

Note that the fermion boundary condition in the time direction is
anti-periodic and in the $SU(2)$ case a change from periodic to antiperiodic
boundary condition is exactly equivalent to flipping the sign of the Polyakov
loop. Figure \ref{fig:denslow_sch0} shows that the gap in the spectrum appears
if the anti-periodic boundary condition and the Polyakov loop combine to give 
an effective anti-periodic boundary condition to the fermions. Moreover, the
gap is roughly equal to the lowest Matsubara frequency of the fermions
at $N_T=4$, which I also indicated in the same Figure. I also checked that 
this also happens for $N_T=6$ (not shown in the Figure). Similar dependence
of the lowest eigenvalue on the boundary condition was observed in the $SU(3)$
case in Ref.\ \cite{Gattringer:2002tg}.

\begin{figure}
\begin{center}
\includegraphics[width=.7\textwidth]{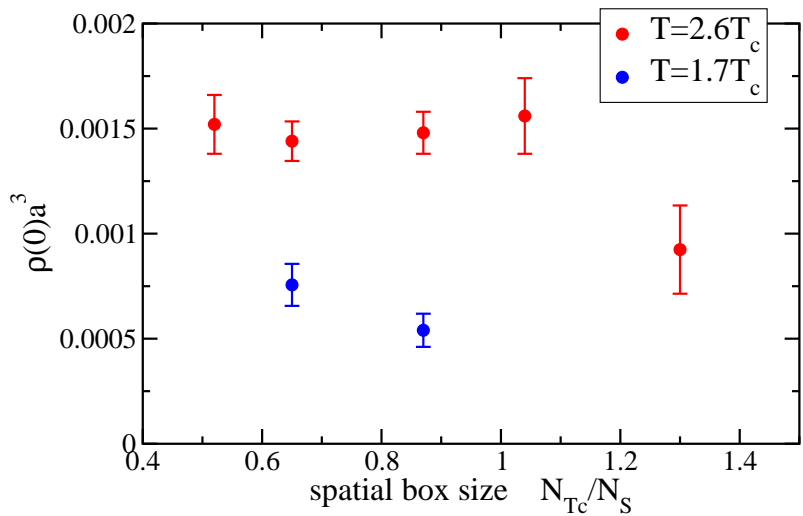}
\caption{$\rho(0)$, the density of modes of the overlap Dirac operator
normalized by the four-volume in spatial boxes of various sizes in the
negative Polyakov loop sector.}
\label{fig:denslow}
\end{center}
\end{figure}

In order to establish the presence of a non-zero chiral condensate in the
negative Polyakov loop sector one should check that the eigenvalue density
around zero scales with the three-volume. In Figure \ref{fig:denslow} I show
the eigenvalue density normalized by the volume in spatial boxes of different
sizes. The size of the lattice in the spatial directions is given in units of
$N_{T_c}$, the length scale corresponding to the critical temperature,
i.e.\ the confinement scale. The eigenvalue density appears to be constant
down to spatial sizes of roughly the confinement scale. This shows that indeed,
the chiral condensate persists even at these temperatures provided the 
Polyakov loop is negative. 

\begin{figure}
\begin{center}
\includegraphics[width=.7\textwidth]{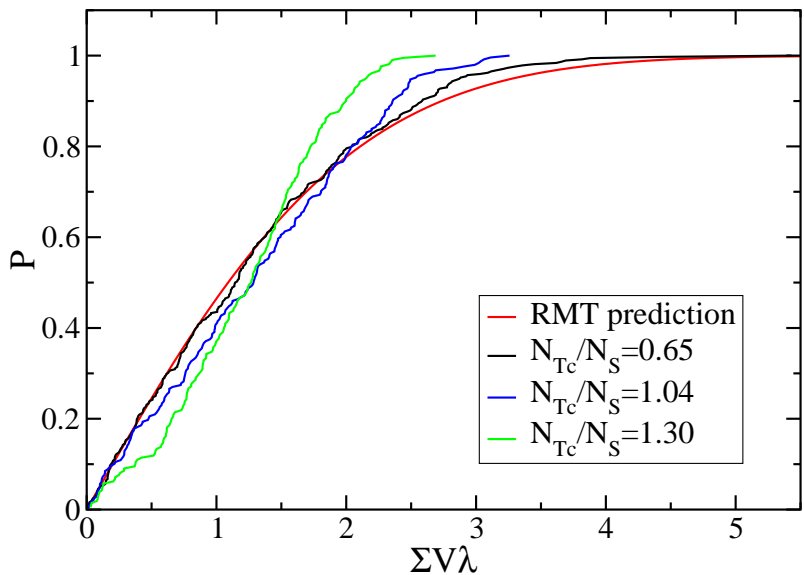}
\caption{The cumulative distribution of the rescaled 
lowest Dirac eigenvalue in the
$Q=0$ topological sector compared to the prediction of random matrix theory
for the chiral orthogonal ensemble.}
\label{fig:rmt}
\end{center}
\end{figure}

This scenario is further supported by comparing the distribution of the lowest
Dirac eigenvalue with the prediction of random matrix theory (RMT). Since at
this high temperature most of the configurations belong to the topological
sector $Q=0$ I do the comparison only in this sector. In Figure \ref{fig:rmt}
I show the cumulative distribution of the lowest Dirac mode compared to the
analytically known RMT prediction for the chiral orthogonal ensemble
\cite{Edwards:1999ra} that is supposed to describe the $SU(2)$ theory. This
comparison involves one adjustable parameter, $\Sigma V$, where $\Sigma$ is
the value of the chiral condensate and $V$ is the volume. $\Sigma V$ is used
to rescale the eigenvalues to fit the eigenvalue density to the universal RMT
curve. If the spatial volume is large enough, the distribution of the rescaled
eigenvalues agrees well with the RMT prediction lending further support to the
presence of a non-zero condensate.

The non-zero Dirac eigenvalue density around $\lambda=0$ and as a consequence
the chiral condensate are usually attributed to mixing instanton anti-instanton
would-be zero modes. Since the topological charge scales with the volume these
would-be zero modes are in principle capable of providing an eigenvalue
density that is also proportional to the volume. There are also speculations
that the chiral condensate in the high temperature phase might also be due to
topological charge fluctuations, in particular calorons. On the other hand, as
the system is heated above the critical temperature, the topological
susceptibility starts to drop sharply. If the understanding of low Dirac
eigenvalues based on topological fluctuations continues to work above the
critical temperature one expects the eigenmode density to drop with the
topological susceptibility. In sharp contrast with that in
Figure \ref{fig:denslow_sch0} the eigenmode density can be seen to increase
with the temperature going up. The opposite behavior of the topological
susceptibility and the eigenmode density strongly suggests that above $T_c$
topological fluctuations become less and less responsible for low eigenmodes
and as the temperature goes further up some other yet unknown mechanism
takes over. 
%This is consistent with the fact that in three-dimensional QCD
%a non-vanishing condensate was found in lattice simulations 
%\cite{Damgaard:1998yv} and as $T\rightarrow\infty$ the four-dimensional 
%system should approach the three-dimensional one by dimensional reduction.

\section{Dynamical fermions: how they select the ``correct'' Polyakov sector?}

We have seen that in the negative Polyakov loop sector there is a chiral
condensate that even increases with the temperature. At this point one could
ask the question how in the real world chiral symmetry is restored above 
$T_c$. The answer is that the fermion determinant breaks the Polyakov loop 
symmetry explicitly. Our experience is that above $T_c$ both in the $SU(2)$
and the $SU(3)$ case the fermion determinant ``favors'' the sector 
where the Polyakov loop lies along the positive real axis. 

This can be qualitatively understood based on the above discussion of how the
gap in the Dirac spectrum is connected to the lowest Matsubara frequency. The
positive real Polyakov loop is exactly the one that combines with the
anti-periodic fermion boundary condition to give the largest possible
effective twist to the fermions in the time direction and the largest first
Matsubara frequency. This in turn results in fewer low modes and a larger
value of the determinant on average.

Is it really only the lowest modes that decide which Polyakov loop sector
is favored by the determinant? Modes higher in the spectrum can also depend on
the Polyakov loop and since there are far more of those, in principle they can
also have a sizeable influence. To decide this I computed all the eigenvalues
of the overlap Dirac operator on small, but physically relevant lattices. If
all the eigenvalues are available in both Polyakov loop sectors one can 
build up the difference in the fermion action starting from the low end of
the spectrum. 

\begin{figure}
\begin{center}
\includegraphics[width=.7\textwidth]{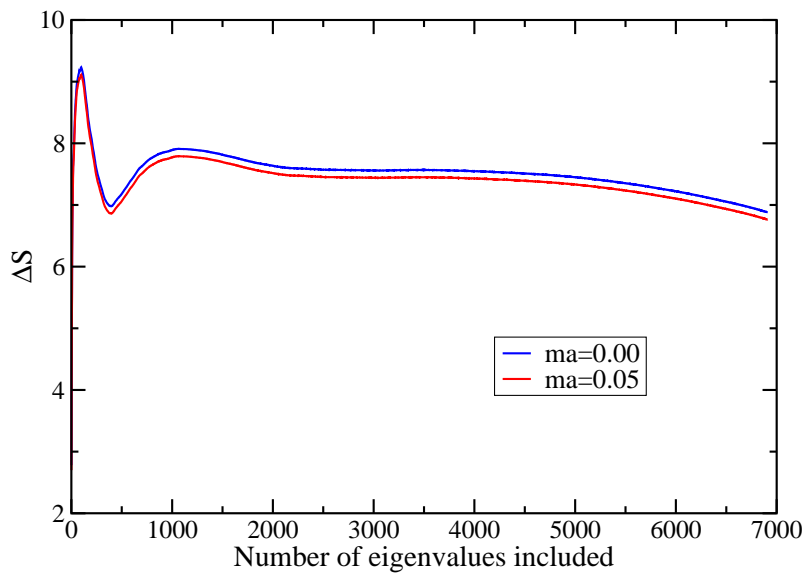}
\caption{The difference in fermion action between periodic and anti-periodic
boundary conditions on a single $6^3\times 4$ configuration at Wilson 
$\beta=2.40$. Flipping the boundary condition is exactly equivalent to 
transforming the configuration to the other Polyakov loop sector. The 
horizontal axis depicts the number of smallest eigenvalues included
when computing the action difference. The two curves both correspond to
a single flavor of fermion, but with different masses.}
\label{fig:actdiff}
\end{center}
\end{figure}

In Figure \ref{fig:actdiff} I plotted how the action difference between the two
sectors depends on the number of eigenvalues included in the determinant. This
was done on a single $6^3\times 4$ configuration at Wilson $\beta=2.40$ by
flipping the boundary condition to reach the other Polyakov loop sector.
Looking at several configurations the pattern seems to be the same. Most of
the action difference comes from a tiny fraction ($<1\%$) of the eigenvalues
at the low end of the spectrum. The same behavior can be confirmed also on
larger configurations and higher $\beta$'s where computing all the eigenvalues
would be prohibitively expensive. In this case I computed only part 
of the spectrum and checked that after including the same small fraction 
of eigenvalues the action difference rapidly stabilizes. Thus it is indeed
the difference in the lowest eigenvalues between the Polyakov sectors that 
is responsible for selecting the ``correct'' Polyakov sector by suppressing 
all other sectors through the fermion determinant. This is consistent with the
fact that eigenvalues higher up in the spectrum are found to be less sensitive
to a change in the boundary condition 
\cite{Bruckmann:2006kx,Synatschke:2008yt}.

\section{Conclusions}

We have seen that in the high temperature phase of the pure $SU(2)$ gauge
theory the density of Dirac eigenmodes around zero strongly depends on the
Polyakov loop. In the quenched theory the Polyakov loop $Z(2)$ symmetry is
spontaneously broken above $T_c$ and I showed that in the $P<0$ sector the low
Dirac mode density and as a consequence the chiral condensate remain non-zero
up to $T=2.6T_c$, the highest temperature considered in the present
study. Since up to this point the eigenvalue density increases with the
temperature, it is very likely that the condensate remains non-zero at
arbitrarily hight temperatures.  I gave further evidence for a non-vanishing
chiral condensate by finding good agreement for the distribution of the lowest
Dirac mode in the $Q=0$ topological sector with the analytically known
prediction of random matrix theory.

In the other Polyakov loop sector above $T_c$, where $P>0$, a gap appears in
the spectrum. I showed that the gap is governed by the lowest Matsubara mode
determined by the effective boundary condition given by the Polyakov loop
combined with the thermal anti-periodic boundary condition. Note that in
contrast to the $SU(2)$ case, if the gauge group is $SU(3)$, the Polyakov loop
cannot exactly cancel the anti-periodic boundary condition. Therefore the gap
is expected to be present in all sectors, but it is still governed by the
lowest free Dirac eigenvalue, as was seen in Ref.\ \cite{Gattringer:2002dv}.

I also showed that contrary to common belief
\cite{Bornyakov:2008bg,Buividovich:2008ip}, a simple model 
based on topological charge fluctuations
is very unlikely to explain the chiral condensate above $T_c$ since the
topological susceptibility decreases, while the chiral condensate increases
with the temperature. It would be interesting to see what other mechanism
could take over the role of the simplest instanton based model. A
possible candidate for that might be dyons \cite{Bornyakov:2008im}.

Finally I showed that by far the most important contribution to the 
fermion determinant is given by a tiny fraction of the lowest eigenvalues.
These are responsible for the fact that dynamical fermions suppress all
Polyakov loop sectors except for the one with the fewest small eigenvalues.
It might be possible to develop this simple qualitative picture into 
a quantitative understanding of how the loss of confinement above $T_c$
implies the restoration of chiral symmetry. Further work in this direction
is in progress.

\end{document}